 \documentclass[preprint2]{aastex}

\begin{document}


 \title{Chemical Abundances in the Secondary Star of\\ the Neutron Star
 Binary \mbox{Centaurus X-4}\footnotetext{Based on observations obtained with
 UVES at VLT Kueyen 8.2 m telescope in programme 65.H-0447}}

 \author{Jonay I. Gonz\'alez Hern\'andez, Rafael Rebolo\altaffilmark{1},\\
 Garik Israelian and Jorge Casares}

 \affil{Instituto de Astrof{\'\i }sica de Canarias, E-38205 La Laguna,
 Tenerife,
 SPAIN: \\jonay@iac.es, rrl@iac.es, gil@iac.es, jcv@iac.es}

 \and

 \author{Keiichi Maeda}
 \affil{Department of Earth Science and Astronomy, Graduate School of Arts
 and Science, University of Tokyo, Meguro-ku, Tokyo 153-8902, JAPAN:
 maeda@esa.c.u-tokyo.ac.jp}

 \and

 \author{Piercarlo Bonifacio and Paolo Molaro}
 \affil{Osservatorio Astronomico di Trieste, via Tiepolo 11, 34131
 Trieste, ITALY\\ bonifaci@ts.astro.it, molaro@ts.astro.it}

 \altaffiltext{1}{Consejo Superior de Investigaciones Cient{\'\i }ficas,
 SPAIN}

 \begin{abstract}

 Using a high resolution spectrum of the secondary star in the neutron star
 binary \mbox{Cen X-4}, we have derived the stellar parameters and veiling
 caused by the accretion disk in a consistent way. We have used a $\chi^{2}$
 minimization procedure to explore a grid of 1\,500\,000 LTE synthetic
 spectra computed for a plausible range of both stellar and veiling
 parameters. Adopting the best model parameters found, we have determined
 atmospheric abundances of Fe, Ca, Ti, Ni and Al. These element
 abundances are super solar ($\mathrm{[Fe/H]}=0.23 \pm 0.10$), but
 only the abundance of Ti and Ni appear to be moderately enhanced
 ($\ge1\sigma$) as compared with the average values of stars of similar iron
 content. These element abundances can be explained if the secondary star captured a
 significant amount of matter ejected from a spherically symmetric supernova
 explosion of a 4 {$M_\odot$} He core progenitor and assuming solar
 abundances as primordial abundances in the secondary star. The kinematic
 properties of the system indicate that the neutron star received a natal
 kick velocity through an aspherical SN and/or an asymmetric neutrino
 emission. The former scenario might be ruled out since our model
 computations cannot produce acceptable fits to the observed abundances. We
 have also examined whether this system could have formed in the Galactic
 halo, and our simulations show that this possibility seems unlikely. We also
 report a new determination of the Li abundance consistent with previous
 studies that is unusually high and close to the cosmic Li abundance in the
 Galactic disk.   

 \end{abstract}

 \keywords{stars:neutron---stars:abundances---stars:individual
 (\mbox{Cen X-4})---stars:X-rays:low-mass---binaries}

 \section{Introduction}

 \begin{figure*}[ht!]
   \centering
   \includegraphics[width=11cm,angle=90]{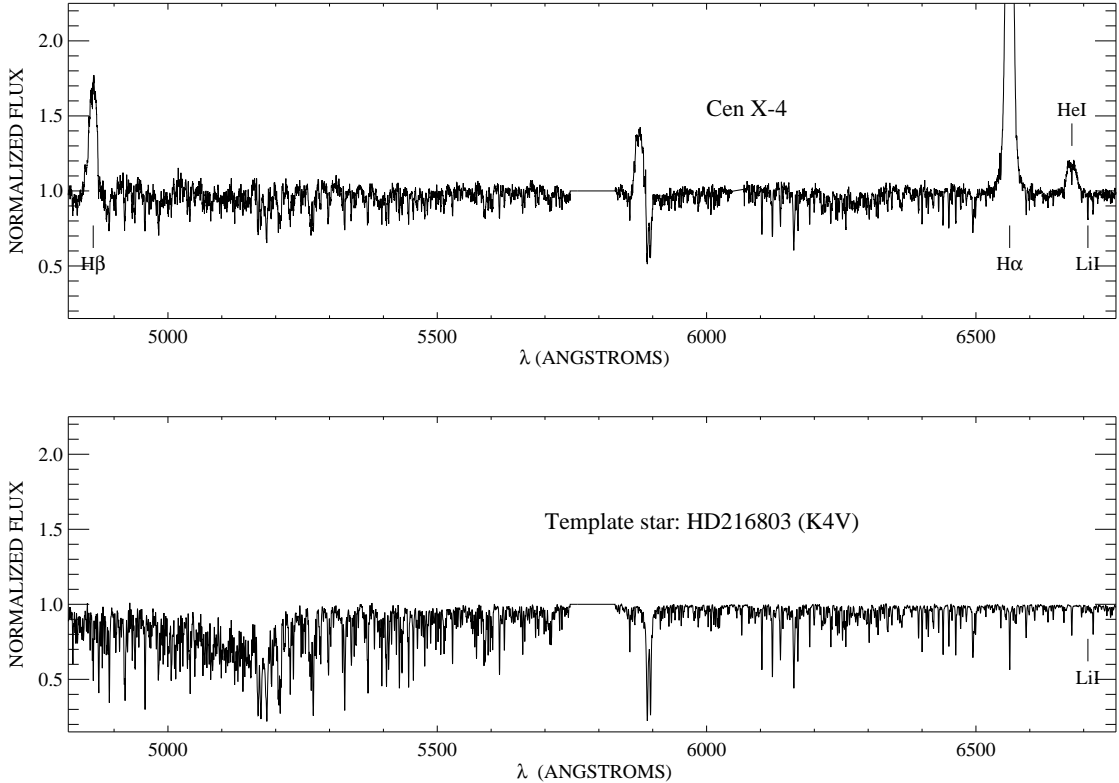}
   \caption{Observed spectrum of the secondary star of \mbox{Cen X-4}
 (top)
   and of a properly broadened template (HD 216803, bottom). \label{fig1}}
 \end{figure*}

 The optical counterpart of the low mass X-ray binary (LMXB) \mbox{Cen
 X-4} (V822 Cen) was firstly identified during an X-ray outburst in 1979
 (Kaluzienski et al.\ 1980), ten years after the X-ray outburst detected by
 the satellite {\it Vela 5B} during an X-ray outburst in 1969 (Conner et al.\
 1969). At that moment, it resembled an $m_{V} \sim 13$ magnitude blue star
 that had increased in brightness by $\sim$6 magnitudes from its pre-outburst
 value (Canizares et al.\ 1980). About three weeks after the 1979 outburst a
 type I X-ray burst was observed, which indicated that the X-ray source is a
 neutron star (Matsuoka et al.\ 1980). Two weeks later the optical
 counterpart had faded to $m_{V} = 18.2$ mag and spectral observations
 revealed the presence of a late-type companion star (K3--K7) in the system
 (van Paradijs et al.\ 1980). Later optical studies determined the orbital
 period, $P = 0.629$ d and a system center-of-mass velocity with respect to
 the Sun of $137 \pm 17$ {${\rm km}\:{\rm s}^{-1}$} (McClintock \& Remillard
 1990). The corresponding mass function was then estimated at $f(M) = 0.23$
 {$M_\odot$}, which further supported the argument for a neutron star primary
 (rather than a black hole, McClintock \& Remillard 1990). More recently,
 from measurements of the orbital inclination and binary mass ratio, the
 compact object mass has been estimated at $M_{\rm NS} =  0.5$--2.1\
 {$M_\odot$} (Shahbaz et al. 1993) and the companion star mass, in the range
 $0.04 < M_2 < 0.58$ {$M_\odot$} (Torres et al.\ 2002). 

 The evolutionary status of this system has been widely discussed in the
 literature. Whereas Chevalier et al.\ (1989) and McClintock \& Remillard
 (1990) had suggested a scenario in which the secondary star was a ``stripped
 giant'' with a very low mass ($M_2\sim0.1$ {$M_\odot$}), Shahbaz et al.\
 (1993) favored the case for a subgiant. In both cases the secondary star
 would be an evolved star that had spent more than 80 per cent of its life
 (i.e., $\sim 10^{9}$--$10^{10}$ yr depending on its initial mass). On the
 contrary, the compact object in this system is a neutron star, probably the
 remnant of the SN explosion of the initial massive component of the binary.
 The ejected material in the SN explosion is composed of heavy nuclei as
 products of the hydrostatic and explosive nucleosynthesis in the massive
 star. 

 Recent studies of chemical abundances of secondary stars in low mass
 X-ray binaries have opened up a new window of information on the later
 stages in the evolution of massive stars. A significant amount of 
 the ejected matter in any supernova explosion that formed the compact
 objects in these systems could be captured by the companion stars. These
 stars could therefore show anomalous abundances as signatures of the
 explosion. Detailed chemical analysis of the atmospheres of the secondary
 stars could constrain many parameters involved in supernova explosion
 models, such as the mass cut, the amount of fall-back matter, any possible
 mixing, and explosion energies and geometries. Indeed, the chemical analysis
 of the secondary stars provides evidence for a supernova event in the
 history of the systems Nova Sco 94 (Israelian et al.\ 1999; Brown et al.\
 2000; Podsiadlowski et al.\  2002) and \mbox{A0620--00} (Gonz\'alez
 Hern\'andez et al.\ 2004). 
 
 In this paper we present the chemical abundance analysis of the secondary
 star of the LMXB \mbox{Cen X-4} and discuss these results in the context
 of possible evolutionary scenarios.

 \section{Observations}

 We obtained 20 spectra of \mbox{Cen X-4} with the UV--Visual Echelle
 Spectrograph (UVES) at the European Southern Observatory (ESO), {\itshape
 Observatorio Cerro Paranal}, using the 8.2 m {Very Large Telescope} (VLT) on
 2000 April 25 and 2000 June 9, covering the spectral regions
 $\lambda\lambda$4800--5800 \AA\ and $\lambda\lambda$5800--6800 \AA\ at
 resolving power $\lambda/\delta\lambda\sim43,000$. Short exposures (718 s)
 were chosen in order to minimize possible smearing of spectral lines
 associated with the radial velocity change during its orbital motion, which we
 estimated to be less than 11 {${\rm km}\:{\rm s}^{-1}$}.

 \begin{deluxetable}{ccc}
 \tabletypesize{\scriptsize}
 \tablecaption{Ranges and steps of model parameters \label{tbl-1}}
 \tablewidth{0pt}
 \tablehead{
 \colhead{Parameter} & \colhead{Range}   & \colhead{Step}
 }
 \startdata
      {$T_{\mathrm{eff}}$}   & {4000  $\to$ 5500 K} & 100 K\\
      {$\log g$}    & {2.5 $\to$ 5}       & 0.1 \\
      {[Fe/H]}      & {$-1.$ $\to$ 1}    & 0.05 \\
      {$f_{4500}$}  & {0 $\to$ 2 }      & 0.10 \\
      {$m_0$}       & {0 $\to$ $-$0.00082}  &$-$0.00002\\
 \enddata

 \end{deluxetable}

 The spectra were reduced in a standard manner using the UVES reduction
 package within the MIDAS environment. The radial velocity for each spectrum
 was obtained from the ephemeris reported in Casares et al.\ (2005, in
 preparation). The individual spectra were corrected for radial velocity
 and combined in order to improve the signal-to-noise ratio. After binning in
 wavelength in steps of 0.21 m{\AA} the final spectrum had a signal-to-noise 
 ratio of 110 in the continuum. This spectrum is shown in Figure 1.

 \begin{figure*}[ht!]
   \centering
   \includegraphics[width=13cm,angle=0]{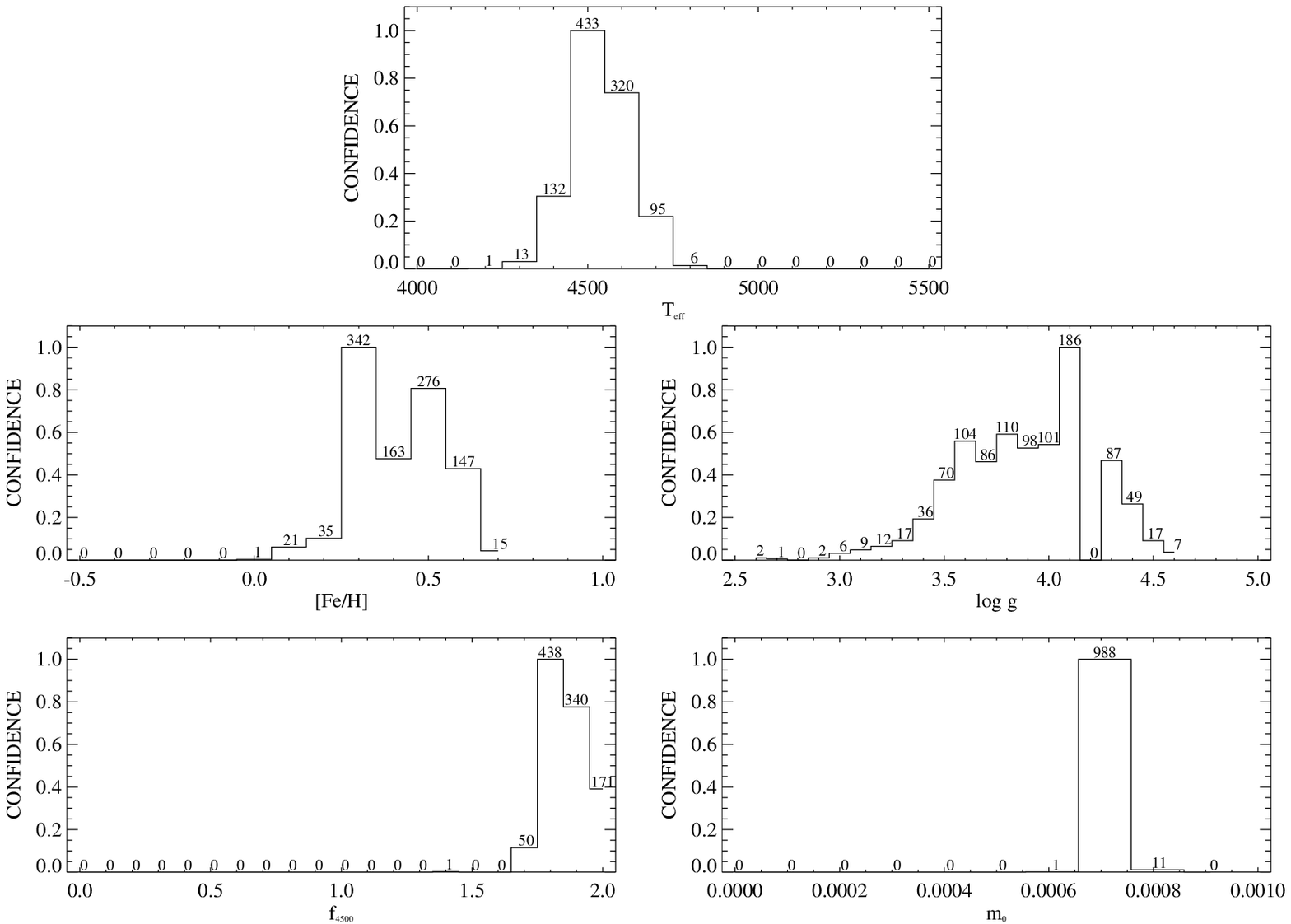}
   \caption{Distributions obtained for each parameter using
   Monte Carlo simulations. The labels at the top of each bin indicate the
   number of simulations consistent with the bin value. The total number
   of simulations was 1000.\label{fig2}}
 \end{figure*}

 \begin{figure*}[ht!]
   \centering
   \includegraphics[width=11cm,angle=90]{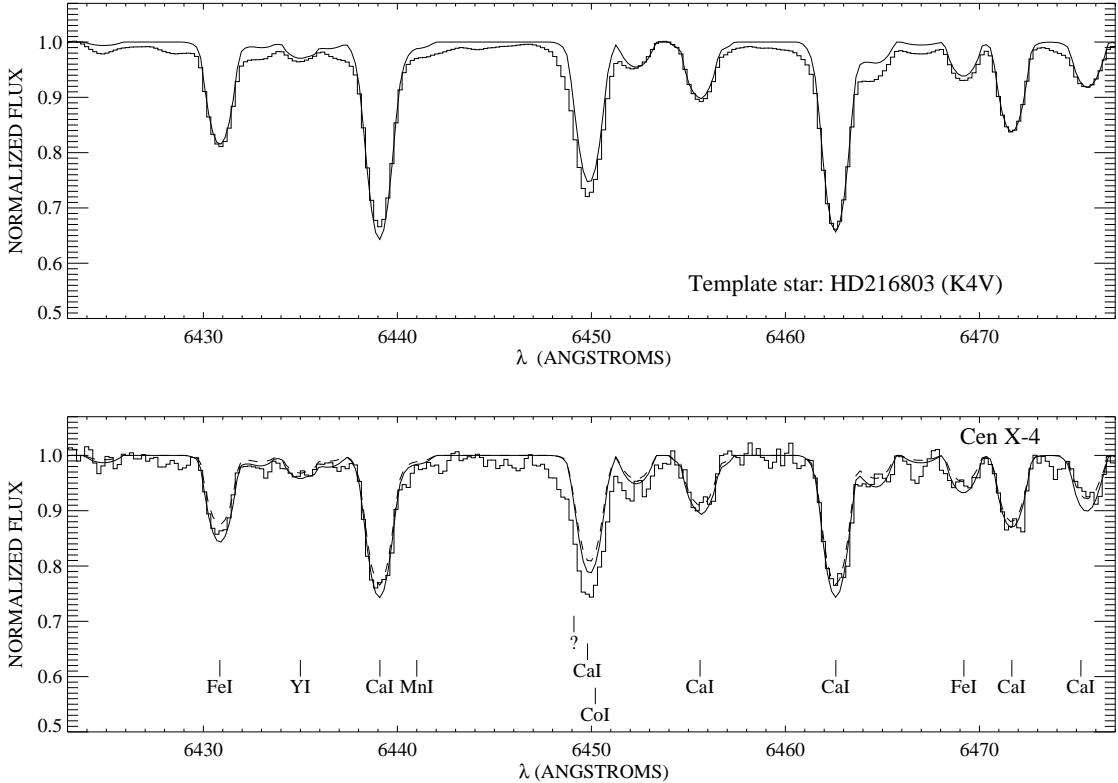}
   \caption{Best synthetic spectral fits to the UVES spectrum of the
   secondary star in the \mbox{Cen X-4} system (bottom panel) and the same
   for a template star (properly broadened) shown for comparison (top panel).
   Synthetic spectra are computed for solar abundances (dashed line) and best
   fit abundance (solid line).\label{fig3}}
 \end{figure*}

 \begin{figure*}[ht!]
   \centering
   \includegraphics[width=11cm,angle=90]{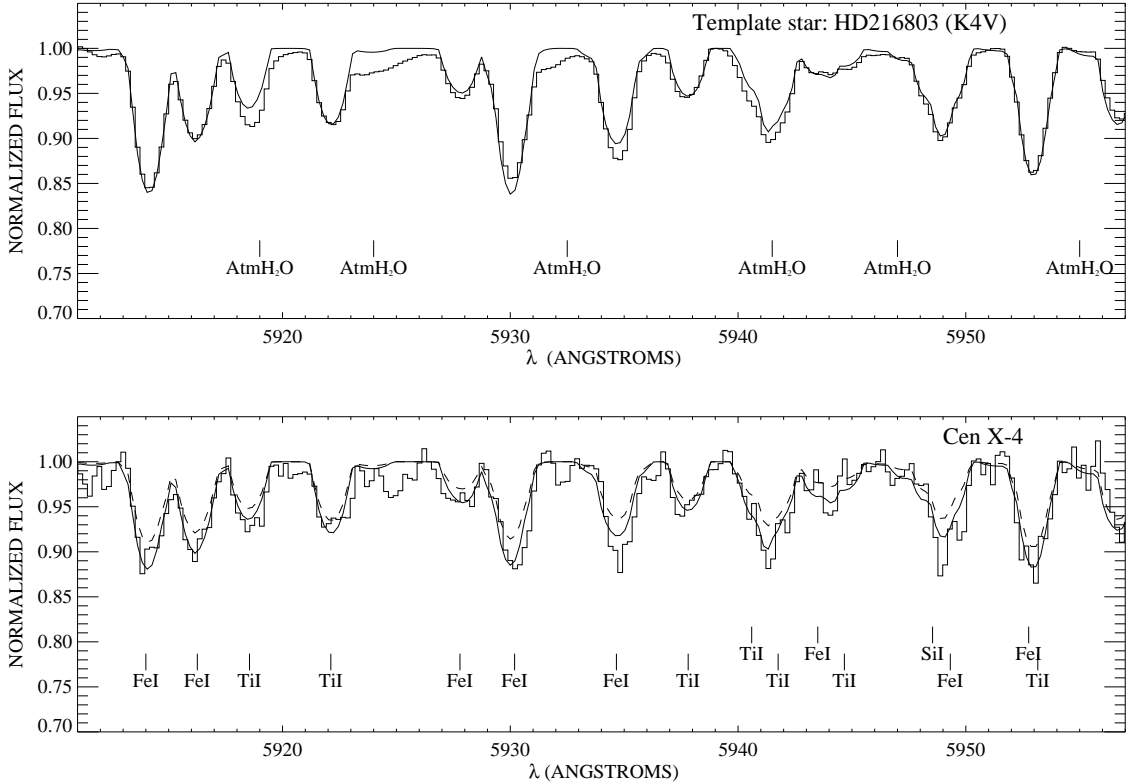}
   \caption{The same as in Figure 3. The spectrum of the template is not
   corrected for telluric lines (atm.\ H$_2$O);  therefore, these lines
   appear broadened in the spectrum (histogram line) and the synthetic
   spectrum does not fit (solid line).\label{fig4}}
 \end{figure*}

 \begin{figure*}[ht!]
   \centering
   \includegraphics[width=11cm,angle=90]{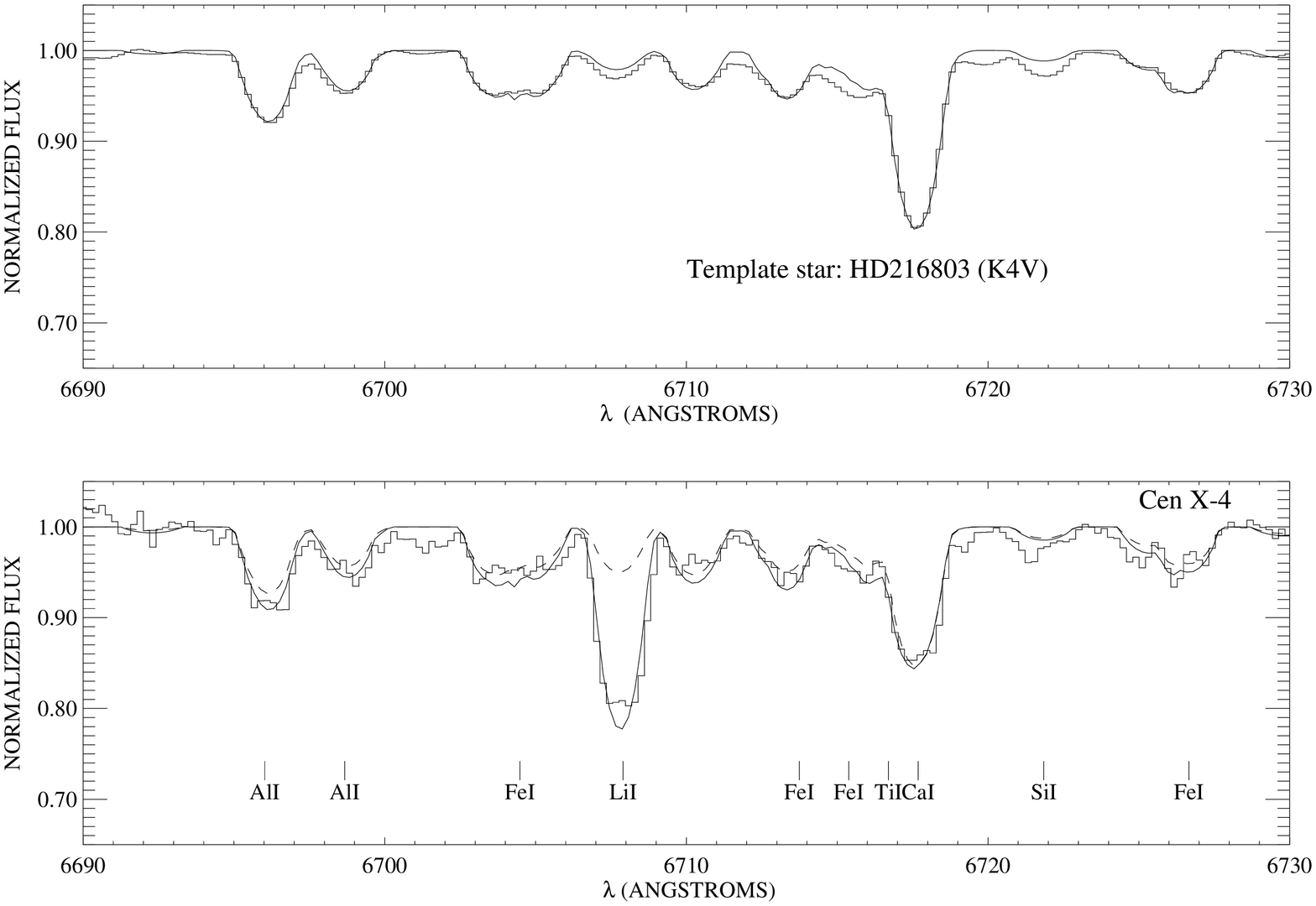}
   \caption{The same as in Figure 3.\label{fig5}}
 \end{figure*}

 \section{Chemical Analysis}

 \subsection{Stellar Parameters}

 The contribution of the emission from the accretion disk to the total
 flux in \mbox{Cen X-4} has been estimated at $\sim$ 25--30\% and 40--50\% in
 the {$V$} and {$B$} bands, respectively, whereas it is at least $80$\% at 
 $\lambda 3800$ \AA (Chevalier et al.\ 1989; Torres et al.\ 2002). As in
 Gonz\'alez Hern\'andez et al.\ (2004), we have obtained the veiling, 
 together with the stellar atmospheric parameters, using synthetic
 spectral fits to the high resolution spectrum of the secondary star in
 \mbox{Cen X-4}. First, moderately strong and relatively unblended lines of
 several elements of interest were identified in the high resolution solar
 flux atlas of Kurucz et al.\ (1984). We selected several spectral features
 containing in total 29 absorption lines of Fe~{\scshape i} with excitation
 potentials between 0.5 and 4.5 eV. In order to compute synthetic spectra for
 these features, we adopted the atomic line data from the Vienna Atomic Line 
 Database (VALD, Piskunov et al. 1995) and used a grid of local thermodynamic
 equilibrium (LTE) models of atmospheres provided by Kurucz (1992, private
 communication). Synthetic spectra were then computed \mbox{using} the LTE
 code MOOG (Sneden 1973). To mi\-ni\-mi\-ze the effects associated with the
 errors in the transition probabilities of atomic lines, we adjusted the
 oscillator strengths, $\log gf$ values of the selected lines until we
 succeeded in re\-pro\-du\-cing the solar atlas of Kurucz et al.\ (1984) with
 solar abundances (Anders \& Grevesse 1989).
 
 \begin{deluxetable}{lcccccc}
 \tabletypesize{\scriptsize}
 \tablecaption{Uncertainties in the abundances of the secondary star in
 \mbox{Cen X-4}\label{tbl-2}}
 \tablewidth{0pt}
 \tablehead{\colhead{Element} & $\mathrm{[E/H]}_{\rm LTE}$ &
 \colhead{$\Delta_{\sigma}$}
 & \colhead{$\Delta_{T_{\mathrm{eff}}}$} &  \colhead{$\Delta_{\log g}$}
 & \colhead{$\Delta_{\rm veiling}$} &\colhead{$\Delta_{\rm
 TOTAL}$\tablenotemark{\dagger}}}
 \startdata
 Al   &  0.30 & 0.12 & 0.05 &  0.05  & 0.10 & 0.17 \\
 Ca   &  0.21 & 0.07 & 0.10 & -0.12  & 0.02 & 0.17 \\
 Ti   &  0.40 & 0.06 & 0.13 &  0.04  & 0.09 & 0.17 \\
 Fe   &  0.23 & 0.09 & 0.02 &  0.02  & 0.03 & 0.10 \\
 Ni   &  0.35 & 0.06 & 0.06 &  0.02  & 0.05 & 0.10 \\
 Li\tablenotemark{\star} & 3.06 & 0.12 & 0.10 & 0.02 & 0.25 & 0.29 \\
 \enddata
 \tablenotetext{\star}{\mbox{Li} abundance is expressed as: \\$$\log
 \epsilon(\mathrm{Li})_{\rm LTE} =
 \log [N(\mathrm{Li})/N(\mathrm{H})]_{\rm LTE} + 12$$}

 \tablenotetext{\dagger}{The total error was calculated using the
 following
 formula: \\$$\Delta_{\rm TOTAL} = \sqrt{\Delta_{\sigma}^2 +
 \Delta_{T_{\mathrm{eff}}}^2 + \Delta_{\log g}^2 + \Delta_{\rm
 veiling}^2}$$}

 \tablecomments{The errors from the dispersion of the best fits to
 different
 features, $\Delta_{\sigma}$, are estimated using the following formula:
 $\Delta_{\sigma} =\sigma/\sqrt{N}$, where $\sigma$ is the standard
 deviation of
 the measurements. Total errors also take into account the uncertainties
 associated with the stellar parameters and the veiling.}

 \end{deluxetable}

 We generated a grid of synthetic spectra in terms of five free parameters,
 three to characterize the star atmospheric model (effective
 tem\-pe\-ra\-tu\-re, $T_{\mathrm{eff}}$, surface gravity, $\log g$, and 
 metallicity, [Fe/H]), and two further parameters to take into account the
 effect of the accretion disk emission on the stellar spectrum. This was
 assumed to be a linear function of wavelength and thus characterized by 
 two parameters: veiling at 4500 {\AA}, $f_{4500} = F^{4500}_{\rm
 disk}/F^{4500}_{\rm sec}$, and the slope, $m_0$. Note
 that the total flux is defined as $F_{\rm total} = F_{\rm disk}+F_{\rm
 sec}$, where $F_{\rm disk}$ and $F_{\rm sec}$ are the flux contributions of
 the disk and the continuum of the secondary star, respectively. These five
 parameters were changed according to the steps and ranges given in Table 1.
 A rotational broadening of 44 {${\rm km}\:{\rm s}^{-1}$}, a limb-darkening
 $\epsilon = 0.65$ and a fixed value for the microturbulence, $\xi = 2$
 {${\rm km}\:{\rm s}^{-1}$}, was adopted. 

 The observed spectrum was compared with each of the 1\,500\,000 synthetic
 spectra in the grid via a $\chi^2$ minimization procedure that provided
 the best model fit. Using a {bootstrap Monte-Carlo method} we defined the
 1$\sigma$ confidence regions for the five free parameters and established
 the most likely values: $T_{\mathrm{eff}} = 4500 \pm 100$ K, $\log g = 3.9
 \pm 0.3$, [Fe/H]$ = 0.4 \pm 0.15$, $f_{4500} = 1.85 \pm 0.10$, and $m_0 =
 -0.00071 \pm 0.00003$. Confidence regions determined using 1000
 realizations are shown in Figure 2. This spectroscopic determination of the
 surface gravity is consistent with derived surface gravities from the mass
 and radius of the secondary star that lie between $\log g \sim 3.6$--4.0.

 \subsection{Stellar Abundances}

 Using the derived stellar parameters we ana\-ly\-zed several spectral
 regions where we have identified various lines of Fe, Ca, Al, Ti, Ni, and
 Li, often blended with Fe lines. Each of these spectral regions was
 carefully normalized using a late type star template (HD 216803) properly
 broadened with the rotational profile used for the secondary star in 
 \mbox{Cen X-4} for comparison. We determined the abundances of these
 elements by comparing the observed spectrum with a grid of synthetic
 spectra through a $\chi^2$ minimization procedure. For these spectral
 syntheses we modified the element abundances while the stellar parameters
 and the suitable veiling factor for each spectral region were kept fixed
 (for further details, see Gonz\'alez Hern\'andez et al.\ 2004).

 In Figs\ 3, 4 and 5 we show several spectral regions where we can see the
 best model synthesis in comparison with the synthesis using solar
 abundances. We also analyzed the broadened spectrum of the template late
 type star and found the abundances of all the elements under study to be
 close to solar (see Bodaghee et al.\ 2003) except the Li abundance which was
 estimated to be $\log \epsilon(\mathrm{Li}) \lesssim 0.7$.

 Abundances for all the elements are listed in Table 2 and referred to the
 solar values adopted from Anders \& Grevesse (1989). We also give
 \mbox{errors}, $\Delta_\sigma$, estimated from the dispersion of the
 elemental abundances obtained from the best fits to the various spectral
 features. Errors associated with uncertainties in effective temperature,
 $\Delta_{T_{\mathrm{eff}}}$, gravity, $\Delta{\log g}$, and veiling,
 $\Delta_{\rm veiling}$, are also listed in Table 2.

 \begin{figure*}[ht!]
 \rotate
   \centering
   \includegraphics[width=13cm,angle=0]{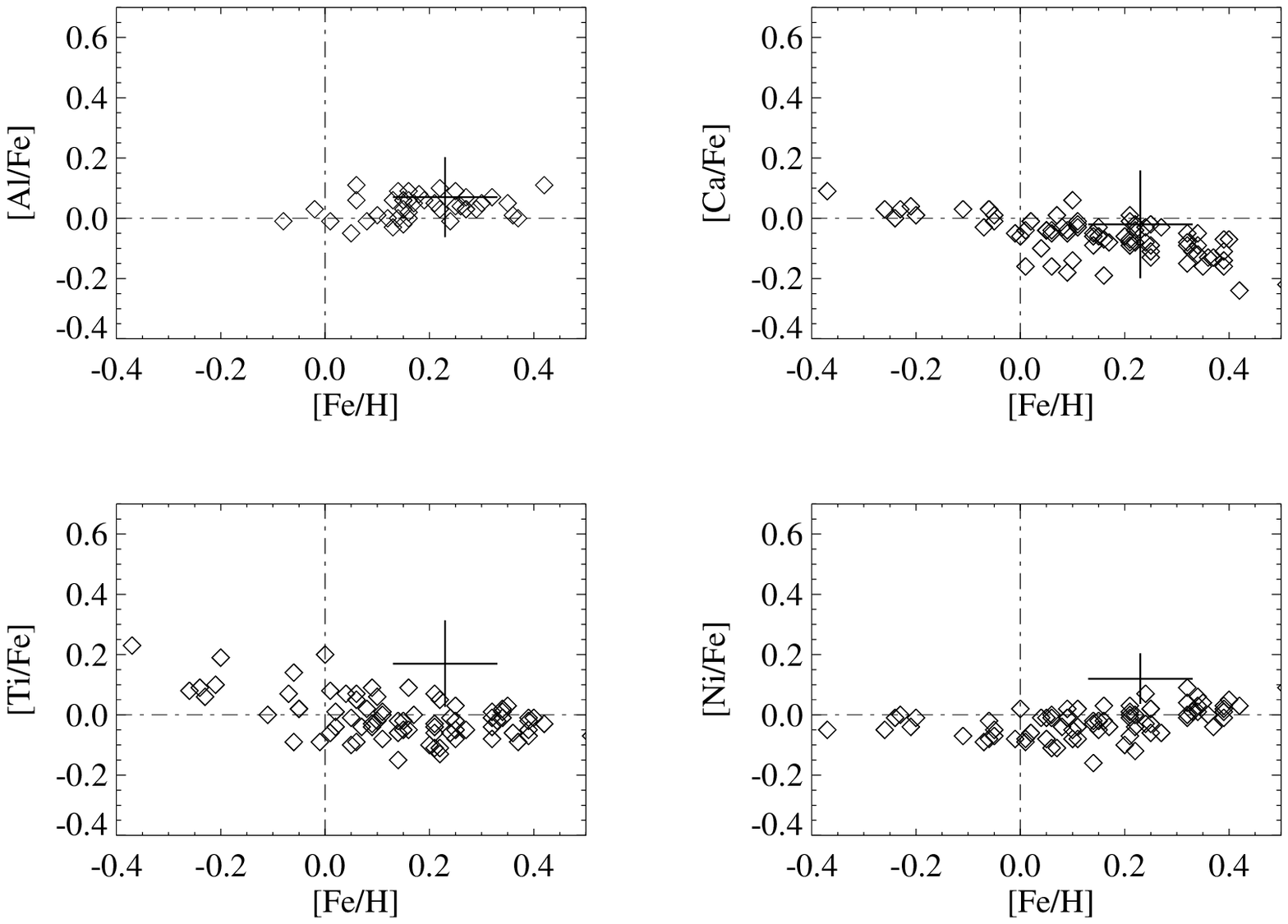}
   \caption{Abundances of the secondary star in \mbox{Cen X-4} (wide
   crosses) in comparison with the abundances in G and K metal-rich dwarf
   stars. Galactic trends of Ca, Ni, and Ti were taken from Bodaghee et al.\
   (2003) while Al from Feltzing \& Gustafsson (1998). The size of the cross
   indicates the size of error. The dashed-dotted lines indicate solar
   abundance values.\label{fig6}}
 \end{figure*}

 The Li feature is particularly strong in our target resulting in an
 abundance of $\log \epsilon(\mathrm{Li})_{\rm LTE}=3.06 \pm 0.29$. We have
 also estimated the non-LTE abundance correction for this element, $$\Delta
 \log \epsilon(\mathrm{Li})= \log \epsilon(\mathrm{Li})_{\rm NLTE}-\log
 \epsilon(\mathrm{Li})_{\rm LTE}$$ from the theoretical LTE and non-LTE
 curves of growth in Pavlenko \& Magazz\`u (1996). By using the derived
 stellar parameters and LTE Li abundance, we obtain $\log
 \epsilon(\mathrm{Li})_{\rm NLTE}= 2.98$ for the secondary star in this
 system. 
 
 The LiI 6708\AA\ resonance doublet may also form in absorption in
 the atmosphere of the accretion disk during outburst decay (Suleimanov \&
 Rebolo 1998). Such lines of Li and possibly of other elements could arise
 if the disk luminosity is roughly 0.1--0.005 the Eddington luminosity.
 However, \mbox{Cen X-4} was observed at quiescence, when its X-ray
 luminosity was close to 10$^{-6}$ the Eddington luminosity (Asai et al.\
 1996). In addition, these lines would be extremely broadened as a
 consequence of the rotation profile of the disk ($\sim500$--1000 
 {${\rm km}\:{\rm s}^{-1}$}). Therefore, we discard that absorption lines
 formed in the accretion disk have any significant influence on the derived
 chemical abundances of the secondary star.  

 In the following section, we will discuss these results in the framework
 of the origin and evolutionary scenario of the \mbox{Cen X-4} system.

 \begin{deluxetable}{lccccc}
 \tabletypesize{\scriptsize}
 \tablecaption{Element abundance ratios in the secondary star in
 \mbox{Cen X-4} and in the comparison sample\label{tbl-3}}
 \tablewidth{0pt}
 \tablehead{\colhead{Element} & \colhead{$\mathrm{[E/Fe]}_{{\rm Cen
 X}-4}$} &
 \colhead{$\Delta^{\tablenotemark{\star}}_{{\rm [E/Fe],Cen X}-4}$} &
 \colhead{$\mathrm{[E/Fe]}_{\rm stars}$} & \colhead{$\sigma_{\rm stars}$}
 &
 \colhead{$\Delta_{\sigma,{\rm stars}}$}}
 \startdata
 Al   &  0.07 & 0.13 &  0.04 & 0.03 & 0.006 \\
 Ca   & $-$0.02 & 0.18 & $-$0.07 & 0.04 & 0.008 \\
 Ti   &  0.17 & 0.14 & $-$0.04 & 0.06 & 0.010 \\
 Ni   &  0.12 & 0.08 & $-$0.02 & 0.05 & 0.010 \\
 \enddata

 \tablenotetext{\star}{Errors in the element abundance ratios
 ($\mathrm{[E/Fe]}$) in the secondary star in Cen X-4.}

 \tablecomments{$\mathrm{[E/Fe]}_{\rm stars}$ indicates the average values
 calculated for stars with iron content in the range $-$0.13 to 0.33
 corresponding to 1$\sigma$ in the $\mathrm{[Fe/H]}$ abundance of the
 secondary star in Cen X-4. Ca, Ti and Ni for the comparison sample have
 been taken from 29 stars in Bodaghee et al.\ (2003) while Al from 28
 stars in Feltzing \& Gustafsson (1998). The uncertainty in the average value
 of element abundance ratios in the comparison sample is obtained as
 $\Delta_{\sigma,{\rm stars}} =\sigma_{\rm stars}/\sqrt{N}$ where
 $\sigma_{\rm stars}$ is the standard deviation of the measurements and $N$,
 the number of stars.}

 \end{deluxetable}

 \section{Discussion}

 \subsection{Heavy Elements}

 The abundances of {\it heavy} elements in the se\-con\-dary star in
 \mbox{Cen X-4} are super solar. In Fig.\ 6 these element abundances relative
 to iron are shown in comparison with the Galactic abundance trends of these
 elements in the relevant metallicity range, taken from Feltzing \&
 Gustafsson (1998) and Bodaghee et al. (2003). The error bars in the element
 abundance ratios ($\mathrm{[E/Fe]}$, see Table 3) take into account how
 individual element abundances depend on the various sources of uncertainty
 when dealing with abundance ratios. The $\mathrm{[Ca/Fe]}$ and
 $\mathrm{[Al/Fe]}$ ratios of the secondary are consistent with abundances of
 stars with similar iron content, while Ni and Ti appear to be moderately
 enhanced. In Table 3 we show the element abundance ratios in the secondary
 star in \mbox{Cen X-4} and average values in stars with iron content in the
 range $-0.13 < \mathrm{[Fe/H]} < 0.33$, the comparison sample, corresponding
 to a 1$\sigma$ uncertainty in the iron abundance of the companion star.
 Whereas $\mathrm{[Ca/Fe]}$ and $\mathrm{[Al/Fe]}$ ratios are consistent with
 average values, the $\mathrm{[Ni/Fe]}$ and $\mathrm{[Ti/Fe]}$ ratios are
 1$\sigma$ higher than the average values of the stars in the comparison
 sample.  

 It has been proposed that LMXBs such as \mbox{Cen X-4} begin their lives as
 wide binaries and evolve through a common-envelope phase where the companion
 spirals into the massive star's envelope (see, for example, Van den Heuvel
 1983; de Kool et al.\ 1987; Nelemans \& Van den Heuvel 2001). The helium
 core of the massive star continues its evolution until dying as a supernova 
 leaving a neutron star or a black hole remnant. Part of the ejected mass
 in the SN explosion could be captured by the companion star, as has been
 found in LMXBs like Nova Sco 94 (Israelian et al.\ 1999) and A0620$-$00
 (Gonz\'alez Hern\'andez et al.\ 2004). We have considered this possibility
 here in the case of the \mbox{Cen X-4} system.

 \subsubsection{Spherical SN explosion}

 A binary system like \mbox{Cen X-4} (with neutron star mass $M_{\rm
 NS}=0.5-2.5$ {$M_\odot$} and the companion star mass in the range
 $M_2=0.04$--0.58 {$M_\odot$}) will survive a spherical SN explosion 
 if the ejected mass, $\Delta M=M_{\rm He}-M_{\rm NS}\le (M_{\rm He}+M_2)/2$.
 Therefore, initial masses for a neutron star and the secondary star of
 $\sim1.4$ {$M_\odot$} and $\sim0.8$--1 {$M_\odot$} respectively would
 imply a mass of the He core before the SN explosion $M_{\rm He}\le3.6$--3.8
 {$M_\odot$}. Using the expressions given by Portegies Zwart et al.\ (1997,
 and references therein), we inferred a mass of the progenitor star, $M_{\rm
 1}\sim 16$ {$M_\odot$}, and a radius $R_{\rm He}\sim 0.8$ {$R_\odot$} for
 the helium core. As in Gonz\'alez Hern\'andez et al.\ (2004), we can also
 estimate the amount of the ejected material in a spherical explosion that
 could be captured by the companion as coming from a central point since the
 pre-SN orbital separation between the binary components would be $a_0 \sim
 1.6$--2 {$R_\odot$} (assuming $M_{\rm He}\sim 4$ {$M_\odot$}, $M_{\rm
 NS}\sim1.5$ {$M_\odot$} and $M_2\sim0.5$--0.8 {$M_\odot$}).  The explosion of
 such a massive primary would have taken  place only $\sim 7\times10^6$ yr after the formation of the system (Brunish
 \& Truran 1982). At that time, the radius of an 0.8 {$M_\odot\; $} secondary
 star would be $\sim 1.2$ {$R_\odot\; $} (D'Antona \& Mazzitelli 1994). If we
 consider a spherically symmetric supernova explosion, taking into account
 the fraction of solid angle subtended by the companion and assuming
 different capture efficiency factors, $f_{\rm capture}$, the amount of mass
 deposited on the secondary can be computed as $$m_{\rm add}=\Delta M (\pi
 R_2^2/4 \pi a_0^2)f_{\rm capture}$$

 \begin{deluxetable}{lccccccccc}
 \centering
 \tabletypesize{\scriptsize}
 \tablecolumns{8}
 \tablecaption{Spherical Supernova Explosion in \mbox{Cen
 X-4}\label{tbl-4}}
 \tablewidth{0pc}
 \tablehead{ & & & & \multicolumn{4}{c}{${\rm [E/H] \:
 EXPECTED}$\tablenotemark{d}}\\
 \cline{5-8}\\
  & & & & \multicolumn{2}{c}{$M_2=0.5$ {$M_\odot$}} &
 \multicolumn{2}{c}{$M_2=0.8$ {$M_\odot$}} \\
 \cline{5-6} \cline{7-8} \\
  & & & & \multicolumn{2}{c}{$M_{\rm cut}${$(M_\odot)$}} &
 \multicolumn{2}{c}{$M_{\rm cut}${$(M_\odot)$}} \\
 \cline{5-6} \cline{7-8} \\
 ELEMENT & ${\rm [E/H]\:\rm OBS.}$\tablenotemark{a} & ${\rm [E/H]\:\rm
 SAMPLE}$\tablenotemark{b} & ${\rm [E/H]}_{\dagger,i}$\tablenotemark{c} &
 1.49 & 1.55 & 1.49 & 1.55
 }
 \startdata
 \multicolumn{8}{c}{${\rm [Fe/H]}_{\dagger,i} = 0$} \\
 \noalign{\smallskip}
 \tableline
 \noalign{\smallskip}
 Al & 0.30 &  0.25 & 0 &  0.27 &  0.47 &  0.27 &  0.48            \\
 Ca & 0.21 &  0.15 & 0 &  0.23 &  0.37 &  0.23 &  0.38            \\
 Ti & 0.40 &  0.18 & 0 &  0.48 &  0.19 &  0.49 &  0.20            \\
 Fe & 0.23 &  0.23 & 0 &  0.23 &  0.23 &  0.23 &  0.23            \\
 Ni & 0.35 &  0.20 & 0 &  0.32 &  0.18 &  0.32 &  0.19            \\
 \noalign{\smallskip}
 \tableline
 \noalign{\smallskip}
 \multicolumn{8}{c}{${\rm [Fe/H]}_{\dagger,i}=-0.4$} \\
 \noalign{\smallskip}
 \tableline
 \noalign{\smallskip}
 Al & 0.30 &  0.25 & \nodata &\nodata&\nodata&\nodata&\nodata \\
 Ca & 0.21 &  0.15 & $-$0.24 &  0.26 &  0.49 &  0.27 &  0.49   \\
 Ti & 0.40 &  0.18 & $-$0.11 &  0.65 &  0.26 &  0.66 &  0.26   \\
 Fe & 0.23 &  0.23 & $-$0.38 &  0.23 &  0.23 &  0.23 &  0.23   \\
 Ni & 0.35 &  0.20 & $-$0.31 &  0.39 &  0.17 &  0.39 &  0.17   \\
 \noalign{\smallskip}
 \tableline
 \noalign{\smallskip}
 \multicolumn{8}{c}{${\rm [Fe/H]}_{\dagger,i}=-0.8$} \\
 \noalign{\smallskip}
 \tableline
 \noalign{\smallskip}
 Al & 0.30 &  0.25 & \nodata &\nodata&\nodata&\nodata&\nodata \\
 Ca & 0.21 &  0.15 & $-$0.71 &  0.24 &  0.50 &  0.23 &  0.51   \\
 Ti & 0.40 &  0.18 & $-$0.63 &  0.67 &  0.17 &  0.67 &  0.17   \\
 Fe & 0.23 &  0.23 & $-$0.78 &  0.23 &  0.23 &  0.23 &  0.23   \\
 Ni & 0.35 &  0.20 & $-$0.65 &  0.41 &  0.14 &  0.41 &  0.15   \\
 \noalign{\smallskip}
 \tableline
 \noalign{\smallskip}
 \multicolumn{8}{c}{${\rm [Fe/H]}_{\dagger,i}=-1.4$} \\
 \noalign{\smallskip}
 \tableline
 \noalign{\smallskip}
 Al & 0.30 &  0.25 & \nodata &\nodata&\nodata&\nodata&\nodata \\
 Ca & 0.21 &  0.15 & $-$1.17 & 0.23 & 0.52 & 0.52 & 0.52        \\
 Ti & 0.40 &  0.18 & $-$1.16 & 0.69 & 0.15 & 0.15 & 0.16        \\
 Fe & 0.23 &  0.23 & $-$1.41 & 0.23 & 0.23 & 0.23 & 0.23        \\
 Ni & 0.35 &  0.20 & $-$1.12 & 0.42 & 0.13 & 0.13 & 0.13        \\
 \enddata
 \tablenotetext{a}{Observed abundances of the
 secondary star in Cen X-4.}

 \tablenotetext{b}{Average abundances in stars of
 the comparison sample (see also Table 4).}

 \tablenotetext{c}{Initial abundances assumed for
 the secondary star in \mbox{Cen X-4} from Allende Prieto et al.\ (2004)
 and Stephens \& Boesgaard (2002)}

 \tablenotetext{d}{Expected abundances of the secondary star}

 \tablecomments{Expected abundances in the secondary atmosphere
 contaminated with nuclesynthetic products of a 16 {$M_\odot$}
 \emph{spherically} symmetric core-collapse explosion model ($M_{\rm He} \sim
 4\ M_\odot$) for two different secondary masses, $M_2$. $M_{\rm cut}$ is the
 mass cut assumed for each model. ${\rm [Fe/H]}_{\dagger,i}$ is the initial
 metallicity of the secondary star. The capture efficiency, $f_{\rm
 capture}$, has been modified in each model until matching the expected with
 the observed iron abundance. Therefore, the mass captured by the companion,
 $m_{\rm add}$, lies between 0.02 ($f_{\rm capture}=0.1$) and 0.11 ($f_{\rm
 capture}=0.5$) for the extreme values of $M_2=0.5$ {$M_\odot$}, $M_{\rm
 cut}=1.49$ {$M_\odot$} and ${\rm [Fe/H]}_{\dagger,i}=0$, and $M_2=0.8$
 {$M_\odot$}, $M_{\rm cut}=1.55$ {$M_\odot$} and ${\rm
 [Fe/H]}_{\dagger,i}=-1.4$} 

 \end{deluxetable}

 In Table 4 we present the expected abundances in the atmosphere of the
 secondary star after contamination from the progenitor of the compact
 object for two plausible secondary masses (0.8 {$M_\odot$} and 0.5
 {$M_\odot$}) at the time of the explosion of the primary. We have used a
 spherically symmetric core-collapse explosion model of He core of mass
 $M_{\rm He}\sim4$ {$M_\odot$} and explosion energy of $\varepsilon =
 10^{51}\:{\rm ergs}$. This energy is deposited promptly in the central
 region of the progenitor core to generate a strong shock wave. The
 subsequent propagation of the shock wave is followed through a
 two-dimensional hydrodynamical code, as has been done for aspherical
 explosion models by Maeda et al.\ (2002). Nucleosynthesis is solved by
 detailed nuclear reaction network calculations including 222 isotopes, 
 from $^{1}$H up to $^{71}$Ge. In the simulations, the directional 
 distribution of the input energy at the mass cut is given in a 
 parametric way to allow asymmetric explosions. One half of the energy is
 deposited as kinetic energy, and the corresponding initial velocities
 ($v_{r}$, $v_{z}$) are imposed as $v_{z} = \alpha z$ and $v_{r} = \beta r$ 
 (here $r$ is the equatorial direction, $z$ the jet direction, and $v_r$ and
 $v_z$ the initially imposed velocity components along each direction). 
 Therefore, the initially imposed velocity is a factor of $\alpha/\beta$
 larger in the jet direction than in the equatorial direction. In the
 spherical symmetric model, the ratio $\alpha/\beta$ is set to be unity,
 therefore the model is similar to canonical 1D explosion calculations (see,
 e.g., Umeda et al. 2000). The details of our explosion calculations can be
 found in Maeda et al. (2002). Our present models are basically the same as
 their models, except for the progenitor's mass and the explosion energy. 
 In our modeling, we have chosen different capture efficiency
 factors, $f_{\rm capture}$, between 0 and 1. However, simulations of type Ia
 SNe suggest that the SN blast wave may induce mass-loss in the secondary
 instead of matter accretion (Marietta et al. 2000) and therefore, high
 values for the capture efficiency may not be appropriated. We also assume that the
 matter captured is well-mixed in the convective envelope of the companion
 star. The mass of the convective zone of the secondary star was fixed at
 $m_{\rm conv} = 0.5$ and $0.6$ {$M_\odot\; $} from the evolutionary tracks
 for a 0.5 and 0.8 {$M_\odot\; $} secondary star, respectively, with an age
 of $\sim 7\times10^6$ yr (D'Antona \& Mazzitelli 1994).  

 The mass cut, $M_{\rm cut}$, that divides the ejecta from the collapsing
 core, is low enough for a significant fraction of iron to be ejected
 and consequently captured by the secondary star. The secondary star could
 therefore have had an initial iron abundance lower than its current observed
 abundance. In order to check this possibility we have assumed different
 initial metallicities for the secondary and primary star in the system,
 including low metallicities (given the high galactic latitude,
 $b\sim24^{\circ}$, of the system that might suggest a halo origin). We have 
 taken the initial element abundances from the galactic trends (Allende
 Prieto et al.\ 2004; Stephens \& Boesgaard 2002) for four given iron
 abundances. We have used nucleosynthetic products of the solar metallicity
 explosion model since the abundance pattern of $\alpha$-nuclei can be
 similar to the solar abundance (Umeda et al.\ 2000). However, the odd $Z$ to
 even $Z$ element ratios become smaller for lower metallicity, and hence we
 have excluded $^{27}$Al from low metallicity models.

 The expected abundances for the solar abundance model presented in Table
 4 can fit very well with the observed abundances in the secondary star for
 lower mass cuts. However, these element abundances are strongly sensitive
 to the mass cut, and therefore for a slighlty higher value of this parameter
 the expected abundances cannot reproduce the observed abundances. This
 strong dependence on the mass cut makes the low metallicity models unable to
 produce acceptable fits to the observed abundances. In addition, note that
 for lower metallicity models the secondary star needs to capture 10--20 per
 cent of the mass in its convective zone from the ejecta (i.e., an
 $f_{\rm capture}\sim 0.2$--0.7) to achieve the current observed abundances.
 Whereas for the solar abundance model by capturing roughly 5\% of its
 convective envelope mass the secondary star can increase sufficiently its
 element abundances (i.e., an $f_{\rm capture}\sim
 0.1$--0.3). According to the chemical abundance pattern in the
 secondary star, we find as very unlikely a halo origin. Although, strictly
 speaking, we may only rule out an origin in the  metal-weak halo. The halo
 is also populated by stars that have been accreted from Milky Way satellites
 and some of these are of high metallicity. The only known satellites that
 are accreting onto the Milky Way and comprise a population as metal-rich as
 the Cen X-4 system are Sagittarius (Ibata et al.\ 1994, Bonifacio et al.\
 2004) and Canis Major (Martin et al.\ 2004, Sbordone et al.\ 2004).
 Sagittarius is in a polar orbit (Ibata et al. 1997) and Canis Major in an
 equatorial orbit (Martin et al. 2004). The present orbit of Cen X-4 (see
 Gonz\'alez Hern\'andez et al. 2005) is not compatible with that of either of
 these galaxies, thus it seems likely that Cen X-4 originated in neither of
 them, but a detailed kinematic study is needed to reject this possibility. 

 \subsubsection{Aspherical SN explosion}

 So far we have been considering a spherically symmetric SN explosion, but
 the kinematic properties of this system might suggest an aspherical SN 
 explosion, as has been proposed for Nova Sco 94 (Brown et al.\ 2000). The
 maximum ejected mass in a spherical SN explosion, $\Delta M \sim 2.4$
 {$M_\odot$}, leads to a system velocity of $\sim$139 ${\rm km}\:{\rm
 s}^{-1}$. Gonz\'alez Hern\'andez et al.\ (2005) measured the proper motion
 of the system and derived a space velocity  of $v_{\rm sys}\gtrsim 300$
 ${\rm km}\:{\rm s}^{-1}$, i.e., considerably larger than what could be
 acquired in the case of a spherical SN explosion. In addition, the Galactic
 space velocity components ($U, V, W$) of the system are significantly
 different from the mean values that characterize the kinematics of stars
 belonging to the thin and the thick disks of the Galaxy. We conclude that
 the nascent neutron star must have received an additional \emph{kick}, most
 probably at birth. Theorists have been trying to explain these kicks, whose
 physical origin remains unclear (see Lai et al.\ 2001 for a review, and
 references therein). A \emph{natal} kick could be generated by an asymmetric
 explosion due to global hydrodynamic perturbations in the supernova core, or
 it could be a result of asymmetric neutrino emission in the presence of
 superstrong magnetic fields ($B\gtrsim 10^{15}\: {\rm G}$) in the
 proto-neutron star and/or non-standard $\nu$ physics. In the latter cases,
 an asymmetry of $\sim$1\% in the neutrino emission is sufficient to account
 for a kick of 300 ${\rm km}\:{\rm s}^{-1}$; therefore, a non-spherically
 symmetric SN explosion is not necessary. Recently, some authors have started
 to explore the asymmetric-collapse kick mechanisms from global hydrodynamic
 pertubations. Two and three dimensional studies predict strong neutron star
 kicks from this mechanism (Burrows \& Hayes 1996; Janka et al.\ 2004) even
 in excess of 1000 km/s. On the contrary, Fryer (2004), on the basis of three
 dimensional supernova simulations, found that even the most extreme
 asymmetric collapses do not produced final neutron star velocities above 200
 ${\rm km}\:{\rm s}^{-1}$. Therefore, an aspherical SN explosion cannot be
 ruled out and might be the explanation for the high space velocity of the
 system. 

 \begin{deluxetable}{lccccccccc}
 \centering
 \tabletypesize{\scriptsize}
 \tablecolumns{8}
 \tablecaption{Aspherical Supernova Explosion in \mbox{Cen
 X-4}\label{tbl-5}}
 \tablewidth{0pc}
 \tablehead{ & & & & \multicolumn{4}{c}{${\rm [E/H] \:
 EXPECTED}$\tablenotemark{d}}\\
 \cline{5-8}\\
  & & & & \multicolumn{2}{c}{$M_2=0.5$ {$M_\odot$}} &
 \multicolumn{2}{c}{$M_2=0.8$ {$M_\odot$}} \\
 \cline{5-6} \cline{7-8} \\
  & & & & \multicolumn{2}{c}{$M_{\rm cut}${$(M_\odot)$}} &
 \multicolumn{2}{c}{$M_{\rm cut}${$(M_\odot)$}} \\
 \cline{5-6} \cline{7-8} \\
 ELEMENT & ${\rm [E/H]\:\rm OBS.}$\tablenotemark{a} & ${\rm [E/H]\:\rm
 SAMPLE}$\tablenotemark{b} & ${\rm [E/H]}_{\dagger,i}$\tablenotemark{c} &
 1.49
 & 1.55 & 1.49 & 1.55}
 \startdata
 \multicolumn{8}{c}{${\rm [Fe/H]}_{\dagger,i} = 0$} \\
 \noalign{\smallskip}
 \tableline
 \noalign{\smallskip}
 Al & 0.30 &  0.25 & 0 &  0.59 &  0.84 &  0.59 &  0.84        \\
 Ca & 0.21 &  0.15 & 0 &  0.43 &  0.24 &  0.43 &  0.24        \\
 Ti & 0.40 &  0.18 & 0 &  0.21 &  0.20 &  0.21 &  0.20        \\
 Fe & 0.23 &  0.23 & 0 &  0.23 &  0.23 &  0.23 &  0.23        \\
 Ni & 0.35 &  0.20 & 0 &  0.18 &  0.22 &  0.18 &  0.22        \\
 \noalign{\smallskip}
 \tableline
 \noalign{\smallskip}
 \multicolumn{8}{c}{${\rm [Fe/H]}_{\dagger,i}=-0.4$} \\
 \noalign{\smallskip}
 \tableline
 \noalign{\smallskip}
 Al & 0.30 &  0.25 & \nodata &\nodata&\nodata&\nodata&\nodata \\
 Ca & 0.21 &  0.15 & $-$0.24 &  0.56 &  0.28 &  0.56 &  0.28   \\
 Ti & 0.40 &  0.18 & $-$0.11 &  0.28 &  0.26 &  0.28 &  0.26   \\
 Fe & 0.23 &  0.23 & $-$0.38 &  0.23 &  0.23 &  0.23 &  0.23   \\
 Ni & 0.35 &  0.20 & $-$0.31 &  0.15 &  0.22 &  0.15 &  0.22   \\
 \noalign{\smallskip}
 \tableline
 \noalign{\smallskip}
 \multicolumn{8}{c}{${\rm [Fe/H]}_{\dagger,i}=-0.8$} \\
 \noalign{\smallskip}
 \tableline
 \noalign{\smallskip}
 Al & 0.30 &  0.25 & \nodata &\nodata&\nodata&\nodata&\nodata \\
 Ca & 0.21 &  0.15 & $-$0.71 & 0.59 & 0.25 & 0.59 & 0.25        \\
 Ti & 0.40 &  0.18 & $-$0.63 & 0.20 & 0.17 & 0.20 & 0.17        \\
 Fe & 0.23 &  0.23 & $-$0.78 & 0.23 & 0.23 & 0.23 & 0.23        \\
 Ni & 0.35 &  0.20 & $-$0.65 & 0.12 & 0.21 & 0.12 & 0.21        \\
 \noalign{\smallskip}
 \tableline
 \noalign{\smallskip}
 \multicolumn{8}{c}{${\rm [Fe/H]}_{\dagger,i}=-1.4$} \\
 \noalign{\smallskip}
 \tableline
 \noalign{\smallskip}
 Al & 0.30 &  0.25 & \nodata &\nodata&\nodata&\nodata&\nodata \\
 Ca & 0.21 &  0.15 & $-$1.17 & 0.61 & 0.25 & 0.61 & 0.25        \\
 Ti & 0.40 &  0.18 & $-$1.16 & 0.19 & 0.15 & 0.19 & 0.15        \\
 Fe & 0.23 &  0.23 & $-$1.41 & 0.23 & 0.23 & 0.23 & 0.23        \\
 Ni & 0.35 &  0.20 & $-$1.12 & 0.11 & 0.20 & 0.11 & 0.20        \\
 \enddata
 \tablenotetext{a}{Observed abundances of the
 secondary star in Cen X-4.}
 \tablenotetext{b}{Average abundances in stars of
 the comparison sample (see also Table 4).}
 \tablenotetext{c}{Initial abundances assumed for the secondary star in
 \mbox{Cen X-4} from Allende Prieto et al.\ (2004) and Stephens \&
 Boesgaard (2002)}
 \tablenotetext{d}{Expected abundances of the secondary star}

 \tablecomments{Expected abundances in the secondary atmosphere
 contaminated with nuclesynthetic products of a 16 {$M_\odot$}
 \emph{non-spherically} symmetric core-collapse explosion model ($M_{\rm He}
 \sim 4\ M_\odot$) for two different secondary masses, $M_2$. $M_{\rm cut}$
 is the mass cut assumed for each model. ${\rm [Fe/H]}_{\dagger,i}$ is the
 initial metallicity of the secondary star. The capture efficiency, $f_{\rm
 capture}$, has been modified in each model until matching that expected with
 the observed iron abundance. Therefore, the mass captured by the companion,
 $m_{\rm add}$, lies between 0.04 ($f_{\rm capture}=0.2$) and 0.22 ($f_{\rm
 capture}=0.8$) for the extreme values of $M_2=0.5$ {$M_\odot$}, $M_{\rm
 cut}=1.49$ {$M_\odot$} and ${\rm [Fe/H]}_{\dagger,i}=0$, and $M_2=0.8$
 {$M_\odot$}, $M_{\rm cut}=1.55$ {$M_\odot$} and ${\rm
 [Fe/H]}_{\dagger,i}=-1.4$} 

 \end{deluxetable}

 Nucleosynthesis in aspherical explosions has been examined by Maeda et
 al.\ (2002), who showed that the chemical composition of the ejecta is
 strongly dependent on direction. In particular, the main effect of the
 asphericity is that elements produced by the strong $\alpha$-rich
 freezeout are greatly enchanced relative to iron (e.g., [Ti/Fe]). This
 effect is more pronounced in the jet direction, where the kinetic energy is
 higher. Therefore, assuming that the jet is perpendicular to the orbital
 plane, i.e., that the secondary star is located in the equatorial plane of
 the helium star, elements like Ti, Ni, and Fe are mainly ejected in the jet 
 direction, while Al, O, Si, S, and Mg are preferentially ejected near the
 equatorial plane. Consequently, we should have found a high Al abundance in
 the companion star in comparison with the other elements studied but this is
 not the case. To compare these effects with our observations of \mbox{Cen
 X-4} quantitatively, we compute an aspherical explosion of a $4 M_{\odot}$
 He core with the explosion energy $10^{51}$ ergs. The asphericity is induced
 by the parametric way (see \S 4.1.1) with the asphericity parameter
 $\alpha/\beta = 2$, i.e., the initial shock velocity at the mass cut is 2
 times larger along the jet direction than in the equatorial direction. In
 table 5 we show the expected element abundances in the secondary star after
 pollution from the ejecta of an aspherical SN explosion of a 4 {$M_\odot$}
 He core. Contrary to the spherical case, Al is dramatically enhanced in all
 the model computations and this effect is even stronger for higher mass
 cuts. Therefore, for the solar abundance model, none of the model
 computations produces acceptable fits, even for lower mass cuts. Note that
 explosive burning in aspherical SN is weak, ejecting little Ti and Ni, in
 the equatorial direction. Thus, the mass cut appropiate to the system, i.e.,
 $\sim 1.4$ {$M_\odot$}, is not deep enough to eject these elements to
 explain the large values of [(Ti, Ni)/H] observed. The mass cut in the
 aspherical case is placed at relatively weak burning region as compared with
 the spherical model. Therefore, the dependence of the model abundance on
 changing the mass cut is different from the spherical model, which also
 makes the aspherical model difficult to fit the observed abundances. 
 
 Note that all the above results of the aspherical explosion rely on the
 assumption that the jet direction is perpendicular to the orbital plane,
 therefore to the equatorial direction of the progenitor star since the star
 may well co-rotate with the binary orbit. The assumption that the jet is
 almost perpendicular to the equatorial direction is justified because recent
 core-collapse theories suggest the jets, if formed,  propagate along the
 progenitor's rotational axis (e.g., Fryer \& Warren 2004 and references
 therein). This is also supported by a {\it Hubble Space Telescope} image of
 SN 1987A (Wang et al. 2002). The ejecta of SN 1987A is elongated toward the
 direction perpendicular to the innermost circumstellar ring, which is
 probably along the equatorial direction (e.g., Chevalier \& Dwarkadas
 1995).  

 In the case of low initial metallicity of the secondary star, higher
 values for the capture efficiency (i.e., $f_{\rm capture}\sim 0.3$--1) are
 required to achieve the observed Fe abundance since this element is mainly
 ejected perpendicularly to the orbital plane. Despite this requirement,
 these simulations cannot fit the observed abundances even for high mass
 cuts.

 In summary, a spherically symmetric SN explosion of a 4 {$M_\odot$} He core
 leaving an NS gives a better fit than an aspherical model. A relatively
 low primary star ($M_1 \lesssim 20$ {$M_\odot$}) seems to leave an NS and to
 explode as a normal SN (with an energy of $\varepsilon = 10^{51}\:{\rm
 ergs}$, e.g., Nomoto et al.\ 2003). Our  result would suggest that a normal 
 SN would be less aspherical than a hypernova, as proposed for the black hole
 in the LMXB Nova Sco 94 (Brown et al.\ 2000). However, we cannot rule out
 that the companion star did not capture a significant amount of matter from
 the ejecta of a SN explosion, and therefore that  the observed abundances
 correspond to its primordial abundances.

  \begin{figure}[ht!]
   \centering
   \includegraphics[width=7.5cm,angle=0]{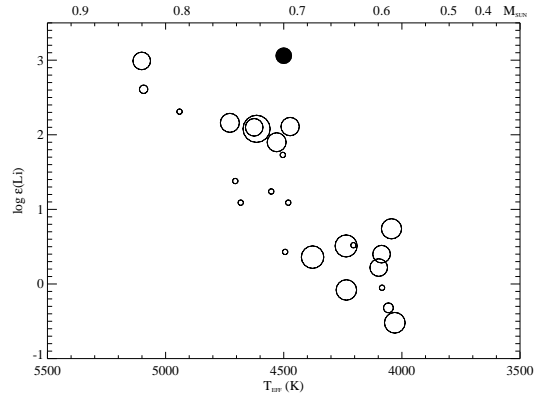}
   \caption{Li abundance of the secondary star in \mbox{Cen X-4} (filled
   circle) in comparison with the abundances of different rotating Pleiades
   dwarf stars versus effective temperature from Garc{\'\i }a L\'opez et
   al.\ (1994). The sizes of the circles are related to $v \sin i$. Stellar
   masses have been assigned following the evolu\-tio\-na\-ry tracks of
   D'Antona \& Mazzitelli (1994) for $10^8$ yr.\label{fig7}}
 \end{figure}

 \subsection{Li abundance}

 The companion star has an unexpectedly high $^{7}$Li abundance close to
 the cosmic value in the Galactic disk (Mart\'\i{}n et al.\ 1994a),
 substantially higher than field late-type main sequence stars of similar
 mass. Convective mixing during pre-main sequence and main sequence evolution
 is expected to produce significant lithium depletion in the atmospheres of
 such stars (see Fig.\ 7), thus a possible explanation for the Li
 over-abundance in the secondary is that the \mbox{Cen X-4} system is younger
 than the Pleiades cluster whose age is $\sim 120$ Myr (Mart{\'\i}n et al.\
 1998; Stauffer et al.\ 1998). One could estimate an upper limit to the age
 of the system under the assumption that there is no mechanism able to enrich
 the atmosphere of the secondary star with freshly synthesized Li nuclei. In
 Fig.\ 8 we show the age of the system for different mass transfer rates and
 initial secondary masses by following the Li depletion in low mass stars
 according to evolutionary tracks of Siess et al.\ (2000). We have adopted
 the NLTE Li abundance and the present value for the initial metallicity of
 the system since a possible metallicity enrichment due to a SN explosion
 would have taken place early enough (at $\sim$7 Myr) and it would make the 
 secondary follow the Li depletion as a metal-rich star. In any case, this
 metallicity effect is only relevant for masses higher than 0.8 {$M_\odot$}.
 Such high Li abundance restricts the age of the system to the range $\sim
 10$--70 Myr. This age is quite short compared to the age assumed 
 in the proposed evolutionary scenarios for the system, where an age of a 
 few times 10$^9$ yr is required (e.g. Chevalier et al. 1989). These age
 constrains are obtained using models that do not consider the impact of high
 rotation on the Li depletion. According to Mart\'\i{}n \& Claret (1996) high
 rotation may inhibit the rate of Li depletion; therefore, the constraints on
 age should be strictly considered as a lower limit. The effect of rotation
 on Li depletion has been measured in Pleiades stars with masses in the range
 0.6--0.8 {$M_\odot$} (see, for example, Garc\'\i{}a L\'opez et al. 1994).
 The kinematic properties of Cen X-4 suggest that it could have been at the
 inner regions of the Galactic disk about 100--200 Myr ago (Gonz\'alez
 Hern\'andez et al.\ 2005). If the neutron star formed at that time as a
 result of an SN explosion, this might explain the high Li abundance of the
 system without postulating a Li production mechanism. 

 For such a young age the effect of the mass transfer in the secondary's
 mass is negligible, and initial secondary masses higher than 0.8
 {$M_\odot$} can therefore be ruled out. Note that the mass transfer rate
 inferred from the observed quiescent X-ray luminosity ($\sim 2.4 \times
 10^{32}$ ${\rm  erg}\:{\rm s}^{-1}$, Asai et al.\ 1996) is very low $\sim
 10^{-6}$ the Eddington rate ($\sim 3 \times 10^{-9}$ $M_\odot\:{\rm
 yr}^{-1}$, Yi \& Narayan 1997). On the other hand, Menou \& McClintock
 (2001) have suggested a mass transfer rate of $\sim$$3 \times 10^{-10}$
 $M_\odot\:{\rm yr}^{-1}$ based on optical--UV emission spectrum of the
 system. 

 \begin{figure}[ht!]
   \centering
   \includegraphics[width=7.5cm,angle=0]{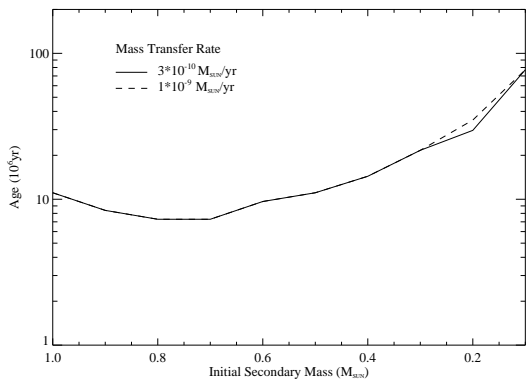}
   \caption{Upper limit to the age of the \mbox{Cen X-4} system according
   to the Li abundance of the secondary star versus initial stellar masses of
   the secondary star, assuming different mass transfer rates. Li depletion 
   has been followed according to the evolu\-tio\-na\-ry tracks of Siess et
   al.\ (2000).\label{fig8}}
 \end{figure}

 High Li abundances appear to be a common property of late-type
 companions in Soft X-ray Transients (SXTs), characterized by strong X-ray
 outbursts (ussually lasting for several months) followed by long quiescent
 periods of $\sim$10--50 yr. Different models have been proposed to explain
 these Li over-abundances in terms of spallation reactions or
 $\alpha$--$\alpha$ reactions. Mart\'\i{}n et al.\ (1994b) proposed that Li
 production could come from nuclear reactions in the accretion disk during
 outburst and being deposited on the companion via a disk wind. They argued
 that the observed abundance must correspond closely to the long term average
 since the depletion time scale ($\tau_D \gtrsim 10^7$ yr) is much longer
 than the recurrence time of outburst ($\sim 30$ yr). Later on, Yi \& Narayan
 (1997) investigated the same scenario, in an advection-dominated accretion
 flow (ADAF), rather than a Shakura--Sunyaev disk not only during outburst
 but also in quiescence, although Li production during outburst may even
 dominate over the production in quiescence. This model can explain the
 observed Li abundance in \mbox{Cen X-4} only if the mass-accretion rate in
 quiescence is $\sim 10^{-3}$ times the Eddington rate and if there is a
 ``propeller'' effect. More recently, Guessoum \& Kazanas (1999) have
 proposed Li production via neutron spallation of CNO elements in the
 secondary star. In this scenario, a flux of neutrons produced in the hot
 accretion flow around the compact object could escape easily and impinge on
 the secondary surface and then produce spallation of heavier elements. A
 short exposure to the above flux, i.e., during outbursts, would suffice to
 produce the observed enhancement. 

 A main sequence K star of mass $\sim$$0.1$--$0.8$ {$M_\odot$} would destroy
 all its primordial Li abundance in $\lesssim 80$ Myr. Yi \& Narayan (1997,
 see also references therein) have discussed in some detail the question of
 Li depletion in the atmosphere of the companion star once it has been
 deposited/produced there, considering the possibility of a steady state,
 in which the enrichment and depletion rates should be equal, as suggested by
 Mart\'\i{}n et al.\ (1994b). Mart\'\i{}n et al.\ (1994a) remarked that
 isotopic ratios $^{7}$Li/$^{6}$Li as low as 5 are expected and hence would
 provide a proof of the spallation scenario. Unfortunately, at such high Li
 abundances there is degeneracy between Li abundance and isotopic ratio and
 consequently, different Li abundances and different isotopic ratios give
 rise to the same line strength. Higher S/N spectra are required to extract
 this information from the 6708 \AA\ Li line although the high rotational
 velocity of the star might make it very difficult. In addition, Guessoum \&
 Kazanas (1999) suggested the possible enhancement of other light elements
 such as Be and B that might be analyzed using UV transitions.

 \section{Conclusions}

 We have obtained a high quality spectrum of the secondary star in
 \mbox{Cen X-4} and derived atmospheric parameters and chemical abundances of
 several elements. We have applied a technique that provides a determination
 of the stellar parameters taking into \mbox{account} any possible veiling
 from the accretion disk. We find $T_{\mathrm{eff}} = 4500 \pm
 100$ K, $\log g = 3.9 \pm 0.3$, and a disk veiling (defined as $F_{\rm
 disk}/F_{\rm total}$) of less than 55 per cent at 5000 {\AA} and
 decreasing towards longer wavelengths.

 The abundances of Fe, Ca, Ti, Al, and Ni are super solar. The abundance
 ratios of each element with respect to Fe were compared with these
 ratios in late-type main sequence metal-rich stars. Moderate anomalies
 for Ni and Ti have been found. A comparison with element yields from
 spherically symmetric supernova explosion models suggests that the secondary
 star could have captured part of the ejecta from a supernova that also
 originated the compact object in \mbox{Cen X-4} if the primordial abundances
 of the secondary star were solar. The observed abundances can be explained
 if a progenitor with a $\sim 4$ {$M_\odot\; $} helium core exploded with a
 mass cut $\sim 1.5$ {$M_\odot$}, such that a significant amount of iron
 could escape from the collapse of the inner layers, and hence the ejecta
 could also have enhanced the iron abundance in the secondary star.

 The high galactic latitude of the system might suggest that the system
 could have originated in the Galactic halo and later enriched by
 nucleosynthetic products ejected in the supernova but our model computations
 for low metallicities show that this possibility is unlikely. In addition, 
 the kinematic properties of this system (Gonz\'alez Hern\'andez et al.\
 2005) suggest that a natal kick was imparted to the neutron star at birth
 due to an asymmetric mass ejection and/or an asymmetry in the neutrino
 emission. We have also inspected a non-spherically symmetric SN explosion
 model but this model cannot produce acceptable fits to the observed
 abundances in the secondary star.

 The Li abundance in the secondary star in \mbox{Cen X-4} is dramatically
 high in comparison with field late-type main sequence stars, possibly
 indicating that this is a young system ($\sim$$10$--$80$ Myr).
 Alternatively, if the system is much older a Li production mechanism is
 needed to balance the large destruction of Li expected in such low-mass
 stars. 

 \section{Acknowledgements}

 This work has made use of the VALD database and IRAF facilities and has
 been partially financed by the Spanish Ministry projects AYA2001-1657,
 AYA2002-03570 and AYA2002-0036.

 \end{document}